\documentclass[onecolumn,amsmath,amssymb]{qm2008_abs}
\usepackage{graphicx}% Include figure files
\usepackage{dcolumn}% Align table columns on decimal point
\usepackage{bm}% bold math
\begin{document}

\title{{\Large First Result of Net-Charge Jet-Correlations from STAR }}% Force line breaks with \\

\bigskip
\bigskip
\author{\large Quan Wang (for the STAR Collaboration)}
\email{wang187@purdue.edu}
\affiliation{Department of Physics, Purdue University, West Lafayette, IN 47907, USA}
\bigskip
\bigskip

\begin{abstract}
\leftskip1.0cm
\rightskip1.0cm
We presented results on azimuthal correlation of net-charge with high $p_T$ trigger particles.
It is found that the net-charge correlation shape is similar to that of total-charge.
On the near-side, the net-charge and total-charge $p_T$ spectra have similar shape and both are harder than the inclusives.
On the away-side, the correlated spectra are not much harder than the inclusives,
and the net-charge/total-charge ratio increases with $p_T$ and is similar to the inclusive ratio.
\end{abstract}

\maketitle

\section{Introduction}
Two novel phenomena have been observed at intermediate transverse momentum ($2< p_T < 5$GeV/$c$) in heavy-ion collisions at RHIC.
First, the azimuthal di-hadron correlation with a high $p_T$ ($p_T > 3$ GeV/$c$) trigger particle
is significantly broadened on the away-side (double-peak structure) in central Au+Au collisions relative to p+p collisions~\cite{bib:broaden}.
Second, the bayon to meson ratio is significantly enhanced~\cite{bib:star}\cite{bib:ratio}.
In order to understand the physics mechanism(s) for these phenomena, we study two-particle azimuthal correlation of net-charge and 
compare to that of total-charge. Since the net-charge is dominated by net-protons, and the total-charge is the sum of 
mesons and bayons, the comparison between net-charge and total-charge azimuthal correlations may help further our understanding of baryon-meson effects at RHIC.

\section{Analysis}
We obtain the raw azimuthal correlations from positively ($h^{+}$) and negatively ($h^{-}$) charged hadrons with a charged high $p_T$ trigger particle ($h$). 
The trigger and associated particle $p_T$ ranges are $3 < p_T < 4$ GeV/$c$ and $1 < p_T < 3$ GeV/$c$, respectively.
Both the trigger and associate particles are restricted to $|\eta|<1$.
Tracking efficiency is corrected for associated particles. The two-particle acceptance effect is corrected for by the event-mixing technique.
The correlations are normalized per trigger particle.
By taking the difference and the sum, we obtain the net-charge ($h\mbox{-}\Delta Q$) and total-charge ($h\mbox{-}Q$) correlations:
\begin{equation}
h\mbox{-}\Delta Q = h\mbox{-}h^{+}-h\mbox{-}h^{-},
\end{equation}
\begin{equation}
h\mbox{-}Q = h\mbox{-}h^{+}+h\mbox{-}h^{-}.
\end{equation}
We obtain the corresponding backgrounds by using the same technique from mixed-events and adding in $v_2$ modulation,
where $v_2$ is presently taken from the charged hadron measurements~\cite{bib:rpv2}\cite{bib:4pv2}. The final correlation signal is given by
\begin{equation}
signal=raw-a*background
\end{equation}
where $a$ is the normalization factor from Zero Yield At 1 radian (ZYA1) method. The systematic uncertainty estimate is done according to Ref.~\cite{bib:broaden}.

\begin{figure}[!ht]
\includegraphics[width=0.7\textwidth]{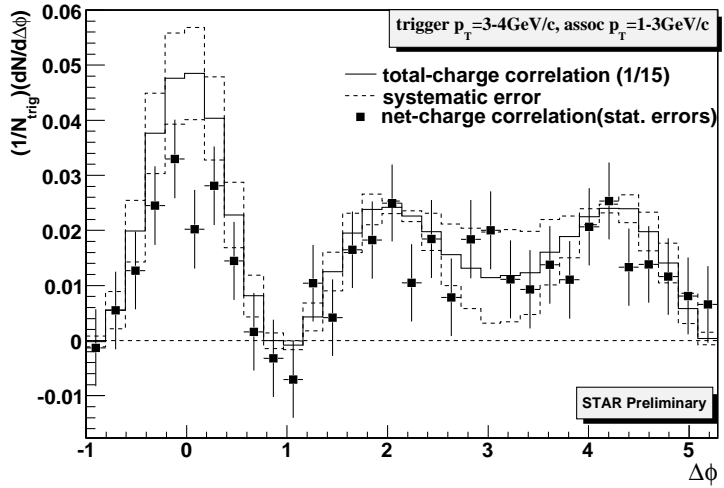}
\caption{\label{fig:corr} Background subtracted azimuthal correlations of net-charge (data points) and total-charge (solid histogram)
in 12\% central Au+Au collisions at 200 GeV. The trigger and associated $p_{T}$ ranges are 3-4 GeV/$c$ and 1-3 GeV/$c$, respectively.
Errors shown for the data points are statistical only.
The total-charge result is scaled by 1/15 to compare with the net-charge result.
The dashed histograms show the systematic errors on the total-charge result,
obtained from background subtraction using the modified reaction-plane $v_2$~\cite{bib:rpv2} and the 4-particle cumulant $v_2$~\cite{bib:4pv2}.}
\end{figure}

\section{Results and Discussion}
Figure~\ref{fig:corr} shows the background subtracted results.
We find the correlation shapes are similar between total-charge and net-charge. The net-charge correlation shows a significant
away-side broadening (double-peak structure), similar to previously observed in total-charge~\cite{bib:broaden}.
We study the $p_T$ dependence of the correlated yields by dividing the data into four associated $p_T$ bins.
In different $p_T$ bins, the correlation shapes are also found to be similar between total-charge and net-charge.

\begin{figure}[!ht]
\includegraphics[width=0.7\textwidth]{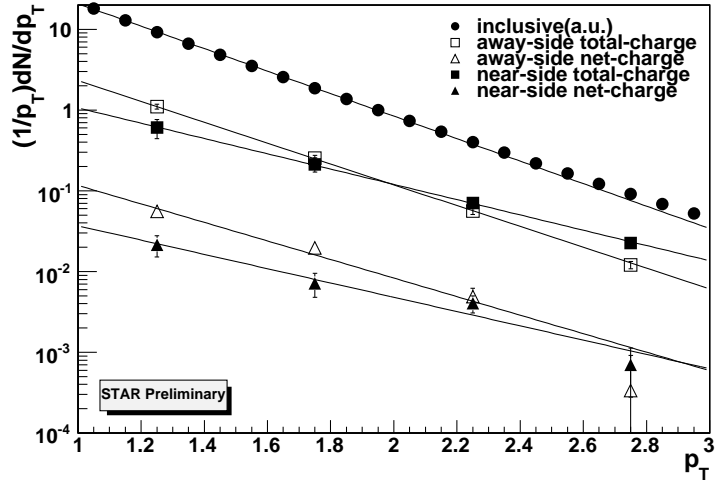}
\caption{
\label{fig:spec} $p_{T}$ spectra of net-charge (triangles) and total-charge (squares) correlated with a high
$p_{T}$ trigger particle ($3<p_{T}<4$ GeV/$c$) on both near-side (filled symbols) and away-side (open symbols) in AuAu central collisions at 200 GeV.
Errors shown are systematic errors. The inclusive (round points) $p_{T}$ spectrum is also shown for comparison.
}
\end{figure}

We obtain associated particle $p_T$ spectra by integrating over the near-side ($|\Delta\phi|<1$) and the away-side ($|\Delta\phi|>1$).
The obtained spectra are shown in Fig.~\ref{fig:spec}.
For comparison, we also show the inclusive charged hadron $p_T$ spectrum (i.e. without requirement of a high $p_T$ trigger particle).
We fit the spectra by an exponential function: $e^{-p_{T}/T}$, where $T$ is the inverse slope parameter. The fit results are listed in Table~\ref{table:T}.

\begin{table}[!h]
\caption{Inverse slope parameter $T$ in GeV/$c$ from exponential fit to the $p_T$ spectra. Errors are statistical only.}\label{table:T}
\begin{tabular}{|c|c|c|}
\hline
 & Near-side & Away-side \\
\hline
net-charge & $0.491\pm0.069$ & $0.378\pm0.032$ \\
\hline
total-charge & $0.459\pm0.003$ & $0.336\pm0.002$ \\
\hline
inclusive & \multicolumn{2}{|c|}{$0.312\pm0.001$} \\
\hline
\end{tabular}
\end{table}

As seen from the extracted values of the inverse slope parameter,
the shapes of net-charge and total-charge correlated spectra on near-side are 
similar, and they are both harder than the inclusive spectrum.
The away-side correlated net-charge spectrum seems to be harder than the total-charge,
and neither is much harder than the inclusive one.

\begin{figure}[!h]
\includegraphics[width=0.7\textwidth]{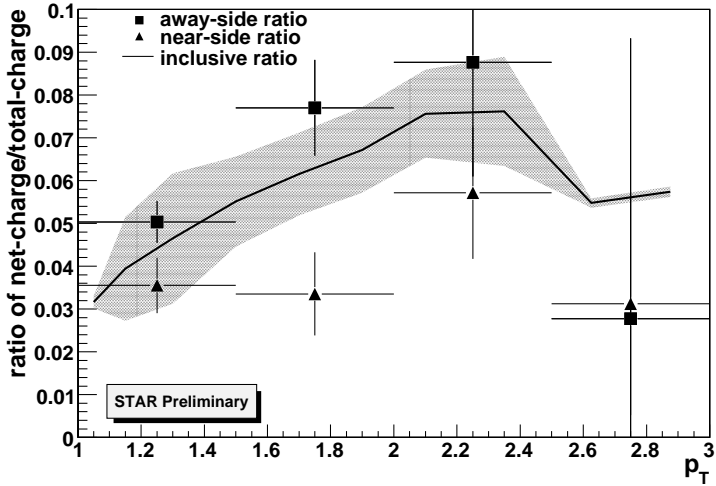}
\caption{
\label{fig:ratio} Ratios of correlated yields of net-charge over total-charge on away-side (squares) and near-side (triangles)
in 12\% central Au+Au collisions at 200 GeV. The trigger $p_{T}$ range is from 3-4 GeV/$c$.
Vertical bars are statistical errors. Horizontal bars are the $p_{T}$ bin size.
The curve shows the inclusive ratio of $\frac{p-\overline p}{\pi^{+} + \pi^{-} + p + \overline p}$,
where the shaded area is statitical uncertainty (the data at $p_T < $2.5 GeV/$c$ are from TOF measurement,
and the $p_T >$2.5 GeV/$c$ data are from TPC measurement)~\cite{bib:star}.
}
\end{figure}

The inclusive bayon to meson ratio is strongly enhanced at intermediate $p_T$ as mentioned earlier.
We want to study the bayon/meson ratio of the correlated hadrons with high $p_T$ trigger particles.
We take the ratio of the $p_T$ spectra on near- and away-side separately.
The obtained results are shown in Fig.~\ref{fig:ratio}, and  compared with the inclusive ratio.
The ratio of the correlated net-charge to total-charge yields on away-side is larger than that on the near-side.
The away-side ratio is similar to the inclusive ratio, with an average difference of $0.34\sigma$.
The near-side ratio seems smaller than the inclusive ratio with an average difference of $0.80\sigma$.

Figure~\ref{fig:ratio} shows that the near-side ratio is independent of $p_T$.
A fit of the near-side ratio to a constant yields $0.037\pm0.005 (stat.)$, with a $\chi^2/ndf = 1.7/3$.
This is to be compared to the inclusive ratio of $0.050\pm0.001$.
On the other hand, the away-side ratio appears to increase with $p_T$, similar to the inclusives.
A constant fit yields $0.055\pm0.004 (stat.)$, with a significantly larger $\chi^2/ndf = 5.4/3$.

\section{Summary}
We have presented results on azimuthal correlation of net-charge with high $p_T$ trigger particles.
The net-charge correlation shape is found to be similar to that of total-charge.
We studied the associate $p_T$ spectra of correlated net-charge, and the net-charge to total-charge ratio.
On the near-side, net-charge and total-charge spectra have similar shape;
the net-charge/total-charge ratio is constant over $p_T$.
However, they are both harder than the inclusives, suggesting jet nature for the correlated particles.
On the away-side, the correlated spectra are not much harder than the inclusives,
suggesting partial equilibration of the correlated hadrons with the medium.
The net-charge/total-charge ratio increases with $p_T$ and is similar to inclusives.

\end{document}